\documentclass[chem-acs, reprint, floatfix, superscriptaddress, showkeys]{revtex4-2}

\usepackage{amsmath}
\usepackage{graphicx}
\usepackage{color}
\usepackage{tabularx}
\usepackage{xr}
\usepackage{array}
\usepackage{upgreek}
\usepackage{moreverb} 

\usepackage{verbatim}

\newcolumntype{Y}{>{\centering\arraybackslash}X}
\newcolumntype{Z}{>{\raggedright\arraybackslash}X}

\makeatletter
\def\@fnsymbol#1{\ifcase#1\or *\or *\or *\else\@ctrerr\fi}
\makeatother

\makeatletter

\begin{document}

\title{Molecularly Thin Polyaramid Nanomechanical Resonators}

\author{Hagen Gress}
\affiliation{Department of Mechanical Engineering, Division of Materials Science and Engineering, and the Photonics Center, Boston University, Boston, Massachusetts 02215, USA \looseness=-1}

\author{Cody L. Ritt}
\affiliation{Department of Chemical and Biological Engineering, University of Colorado Boulder, Boulder, Colorado 80309, USA \looseness=-1}

\author{Inal Shomakhov}
\affiliation{Department of Mechanical Engineering, Division of Materials Science and Engineering, and the Photonics Center, Boston University, Boston, Massachusetts 02215, USA \looseness=-1}

\author{Kaan Altmisdort}
\affiliation{Department of Mechanical Engineering, Division of Materials Science and Engineering, and the Photonics Center, Boston University, Boston, Massachusetts 02215, USA \looseness=-1}

\author{Michelle Quien}
\affiliation{Department of Chemical Engineering, Massachusetts Institute of Technology, Cambridge, Massachusetts 02139, USA \looseness=-1}

\author{Zitang Wei}
\affiliation{Department of Chemical Engineering, Massachusetts Institute of Technology, Cambridge, Massachusetts 02139, USA \looseness=-1}

\author{John R. Lawall}
\affiliation{National Institute of Standards and Technology, Gaithersburg, Maryland 20899, USA \looseness=-1}

\author{Narasimha Boddeti}
\affiliation{Washington State University, School of Mechanical and Materials Engineering, Pullman, Washington 99163, USA \looseness=-1}

\author{Michael S. Strano}
\thanks{Email: strano@mit.edu}
\affiliation{Department of Chemical Engineering, Massachusetts Institute of Technology, Cambridge, Massachusetts 02139, USA \looseness=-1}

\author{J. Scott Bunch}
\thanks{Email: bunch@bu.edu}
\affiliation{Department of Mechanical Engineering, Division of Materials Science and Engineering, and the Photonics Center, Boston University, Boston, Massachusetts 02215, USA \looseness=-1}

\author{Kamil L. Ekinci}
\thanks{Email: ekinci@bu.edu}
\affiliation{Department of Mechanical Engineering, Division of Materials Science and Engineering, and the Photonics Center, Boston University, Boston, Massachusetts 02215, USA \looseness=-1}

\date{\today}

\begin{abstract}
\noindent\textbf{Abstract:} Two-dimensional polyaramids exhibit strong hydrogen bonding to create molecularly thin nanosheets analogous to graphene. Here, we report the first nanomechanical resonators made out of a two-dimensional polyaramid, 2DPA-1, with thicknesses as small as 8~nm. To fabricate these molecular-scale resonators, we transferred  nanofilms of 2DPA-1 onto  chips with previously-etched arrays of circular microwells. We then characterized the thermal resonances of these resonators under different conditions. When there is no residual gas inside the 2DPA-1-covered microwells, the  eigenfrequencies are well-described by a tensioned plate theory, providing the Young's modulus and tension of the 2DPA-1 nanofilms. With gas present, the nanofilms bulge up and mechanical resonances are modified due to the adhesion, bulging and slack present in the system. The fabrication and mechanical characterization of these first 2DPA-1 nanomechanical resonators represent a convincing path toward molecular-scale polymeric NEMS with high mechanical strength, low density, and synthetic processability.
\end{abstract}



\maketitle

\begin{figure*}
    \includegraphics[width=7in]{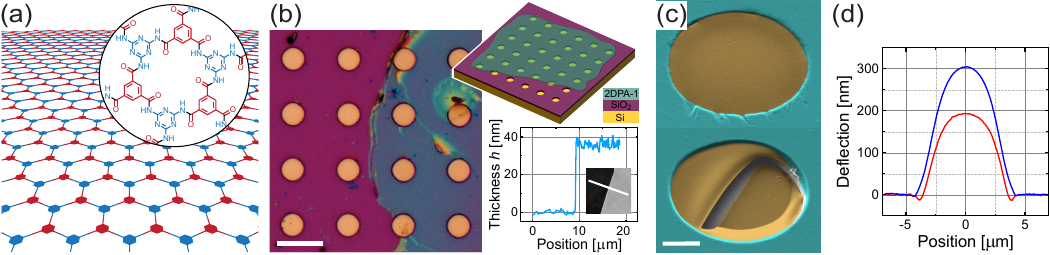}
    \caption{(a) Illustration of an ideal 2DPA-1 monolayer with the molecular structure shown in the inset. (b) Optical microscope image of a typical sample showing the edge of the 2DPA-1 film. Microwells are etched into the $\rm SiO_2$ layer on top of a Si substrate. A 2DPA-1 film is then transferred onto the chip through a wet process to create suspended membranes. The eight microwells on the left are not covered by the film. The scale bar is $20~\rm\upmu m$. The top right inset shows an illustration of the sample. The bottom right inset shows an AFM line scan across the edge of a 35-nm-thick film, along the white line in the corresponding image. (c) SEM images of an intact (top) and a ruptured (bottom) 35-nm-thick membrane. The suspended region is false-colored in orange, and the rest is colored in blue. The scale bar is $2~\rm\upmu m$. (d) AFM line scans through the center of 35-nm-thick circular membranes of radius  $4.25~\rm\upmu m$ at atmospheric pressure. The data were taken within one hour after the membranes were removed from a high pressure chamber filled with nitrogen at $50~\rm kPa$ (red) and $100~\rm kPa$ (blue) above atmospheric pressure. Note the region adhered on the wall in the red curve.}
    \label{Figure_1}
\end{figure*}

Nanoelectromechanical Systems (NEMS) or nanomechanical resonators are evolving both in functionality and form since the first NEMS devices  etched out of silicon in the 1990s \cite{cleland1996fabrication,carr1997fabrication}. An overarching research theme in the field has been the exploration of different materials for NEMS. New materials with extraordinary  properties have allowed for superior resonator parameters as well as the reduction of  resonator linear dimensions. NEMS made out of one- and two-dimensional (2D) nanomaterials --- with unique mechanical, electronic, chemical, and optical properties --- have opened the door for devices with linear dimensions well below the limits of lithography and functionalities well beyond those of semiconductors and metals \cite{yildirim2020towards,xu2022nanomechanical,ferrari2023nanoelectromechanical}.  So far,  2D NEMS have been fabricated from a variety of 2D crystalline  nanomaterials, such as graphene \cite{bunch2007electromechanical}, hexagonal boron nitride \cite{cartamil2017mechanical},  transition metal dichalcogenides \cite{lee2013high,liu2015optical}, MXenes \cite{xu2022electrically,ye2021ultrawide}, and nanoparticles \cite{kanjanaboos2013self,markutsya2005freely,wang2019adaptive}.
Various polymeric materials have also been explored as mechanical resonators \cite{mcfarland2004injection,bunyan2017mechanical,gaitas2006experimental,yoon2019hydrogel,zhang2005electrostatically,adiyan2019shape} --- although these structures have typically remained far from molecular scale thicknesses. Converging these two approaches, i.e., creating molecular scale polymeric NEMS analogous to 2D crystalline NEMS, would open up a powerful new direction in the NEMS field.

Recently, a 2D polyaramid, called 2DPA-1 \cite{zeng2022irreversible}, was synthesized. Molecular-scale disks of 2DPA-1 form by polycondensation of melamine and trimesoyl chloride in solution and assemble into aligned layers upon spin coating (Figure~\ref{Figure_1}a). Each molecular layer is approximately $\sim 3.7~{\rm\r{A}} (=370~\rm pm)$ thick and stacks via hydrogen bonding to form near-molecular-thickness films with an rms surface roughness of $500~\rm pm$ \cite{zeng2022irreversible, ritt2024molecularly}. Despite lacking crystallinity, these films possess mechanical and gas barrier properties that are closer to 2D crystalline nanomaterials, such as graphene, than conventional polymers \cite{ritt2024molecularly}, and can be transferred onto substrates just like graphene. In short, 2DPA-1 combines material properties of conventional 1D polymers and 2D inorganic crystalline nanomaterials, such as graphene; it also opens up a new class of 2D nanomaterials that can be tailored to specific applications using the tools of organic chemistry \cite{zhuang2015two,ren2024recent}. Such tunability --- for example, in gas permeability or in the incorporation of functional groups --- could enable a wide range of future technologies requiring high sensitivity and selectivity. This motivates an exploration of various nanoscale devices that can be fabricated from these new 2D polymeric materials.  To this end, we report the fabrication and measurement of the first nanomechanical resonators made out of 2DPA-1. In addition to demonstrating nanomechanical resonances,  we determine the Young's modulus from the measured resonances; we also develop a mechanical model for a membrane resonator partially adhered to walls and estimate the adhesion energy (between the polymer and the substrate) from  resonance frequencies of membranes delaminating from the substrate walls.

Figure~\ref{Figure_1}b and \ref{Figure_1}c respectively show optical and scanning electron microscope (SEM) images of our nanomechanical drum resonators. Here, 2DPA-1 films are transferred onto  $\rm SiO_2$ substrates via wet transfer \cite{zeng2022irreversible}. The substrates have arrays of etched microwells with depths of $g=960~\rm nm$ and  radii $R$ of either $4.25~\rm\upmu m$ or $2.75~\rm\upmu m$. In Figure~\ref{Figure_1}b, the 2DPA-1 film partially covers the chip. The right insets show  an illustration (top) and  an AFM line scan (bottom) across the edge of the film. The SEM images in Figure~\ref{Figure_1}c show an intact (top) and a broken (bottom)  membrane with $R=4.25~\rm\upmu m$ and thickness $h=35~\rm nm$.

The samples in this study are measured in two distinct pressure states: i) nearly flat with very low gas pressure  on either side, and ii) bulged with finite  gas pressure $p_{in}$ inside the microwell and very low pressure $p_{ext}$ outside. Although the membranes are highly impermeable \cite{ritt2024molecularly}, gas transport into and out-of the microwell can take place through the interface between the polymer film and the substrate \cite{lee2019sealing, manzanares2020improved, ambjorner2023thermal}. When the samples are placed in a high pressure gas chamber ($p_{ext}=120-300~\rm kPa$), the gas leaks into the microwell through the interface \cite{ritt2024molecularly}. Once the sample is transferred to a low pressure environment (vacuum chamber or atmosphere), the pressure difference, $\Delta p=p_{in}-p_{ext}$, across the membrane causes it to initially bulge up. Figure~\ref{Figure_1}d shows two AFM line scans of membranes at atmospheric pressure shortly after removal from a high pressure chamber. When brought to atmospheric pressure, bulged membranes gradually deflate on timescales ranging from minutes to several years \cite{ritt2024molecularly}. If the samples are put in a low pressure chamber (i.e., $p_{ext} \approx 0$), the gas inside the microwell eventually leaks out, allowing the membrane to deflate to a nearly flat state. When $\Delta p=0$, the membrane may be partially adhered to the inside wall of the microwell (Figure~\ref{Figure_1}c) in a stretched configuration with tension. If $\Delta p$ is increased, the membrane starts to delaminate from the wall (Figure~\ref{Figure_1}d), which typically introduces some slack \cite{bunch2008impermeable,calis2023adhesion,cao2024innovative}. We  analyze both states in systematic experiments below.


\begin{figure*}
    \includegraphics[width=7in]{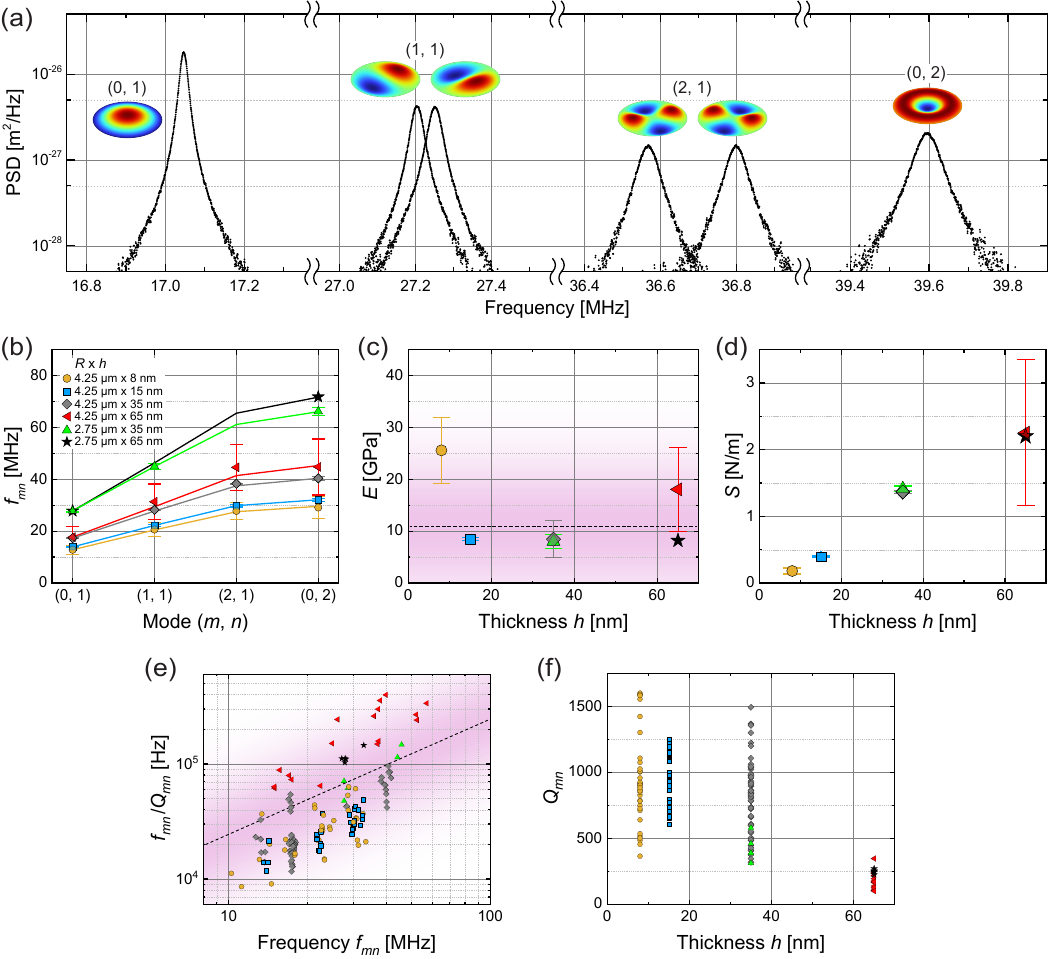}
    \caption{(a) PSD of displacement fluctuations of the first four modes of a membrane with $R=4.25~\rm \upmu m$ and $h=35~\rm nm$. (b) Resonance frequencies $f_{mn}$ for membranes with indicated  $R$ and $h$. The lines show the theoretical frequencies of the first four modes calculated using Eq.~(\ref{eq:f_plate_tension}). (c) Young's modulus $E$ and (d) tension $S$ of the membranes shown in (b) as a function of thickness $h$. The dashed line and the shading respectively indicate the mean and spread of data with the  Young's modulus being $E=11.2 \pm 8.8~\rm GPa$. (e) Dissipation constants $f_{mn}\over Q_{mn}$ as a function of frequency for all measured modes. The shading highlights that ${f_{mn}\over Q_{mn}}\propto f_{mn}$, and the dashed line is a linear fit through the origin. (f) Quality factors $Q_{mn}$ of all measured modes as a function of 2DPA-1 thickness.} 
    \label{Figure_2}
\end{figure*}

We first measure the Brownian motion, i.e., thermal displacement fluctuations, of nearly flat membranes in vacuum. We achieve the flat  state by placing the samples into a chamber  maintained at a vacuum of $\sim 10^{-7}~\rm Torr~(10^{-5}~Pa)$ by an ion pump, which results in $p_{in} \approx p_{ext} \sim 10^{-7}~\rm Torr$. The membranes display detectable thermal resonances for the first few eigenmodes ($m$,~$n$) due to the relatively low dissipation with  resonance frequencies $f_{mn}$ and quality factors $Q_{mn}$. We use a path-stabilized homodyne Michelson interferometer to measure the thermal resonances at an antinode of each mode \cite{gress2023multimode}. Details of the measurement process can be found in the Supporting Information file. Figure~\ref{Figure_2}a shows the power spectral density (PSD) of the displacement fluctuations of the first four eigenmodes of a membrane with $R=4.25~\rm\upmu m$ and $h=35~\rm nm$.  For $m\neq 0$, two nominally degenerate modes exist. If the resonator geometry deviates from an ideal circle, the resonance frequencies of these modes can split --- as noticeable for the  modes $(1,1)$ and $(2,1)$ in Figure~\ref{Figure_2}a. The split modes can be clearly resolved by changing the angular position of the laser spot on the membrane. We fit each peak with a Lorentzian curve to determine $f_{mn}$ and $Q_{mn}$. 

To analyze the eigenfrequencies, we turn to the equation for the free transverse vibrations of a uniform circular plate under tension with radius $R$, thickness $h$, density $\rho$, tension $S$, and bending stiffness $D={Eh^3 \over{12(1-\nu^2)}}$, where  $E$ is the Young's modulus and $\nu$ is the Poisson's ratio:
\begin{equation}
    \rho h{{{\partial ^2}W} \over {\partial {t^2}}} + {D \over {{R^4}}}{\nabla ^4}W - {{S} \over {{R^2}}}{\nabla ^2}W = 0.
    \label{eq:eom}
\end{equation}
Here, ${\nabla ^2}$ is the Laplacian operator in cylindrical coordinates; ${\nabla^4}=\left({\nabla ^2}\right)^2$; and $W(r,\theta)$ is the displacement at $(r,\theta)$. The coordinate $r$ has been normalized with $R$ such that $0 \le r \le 1$. A dimensionless tension parameter $U={{SR^2} \over {D}}={{12(1-\nu^2)SR^2}\over{Eh^3}}$ emerges from Eq.~(\ref{eq:eom}) with the plate and membrane limits  for $U\to 0$ and $U\to \infty$, respectively. Under clamped boundary conditions, Eq.~(\ref{eq:f_plate_tension}) yields the resonance frequency of each eigenmode as  
\cite{wah1962vibration}
\begin{equation}
    f_{mn} = \frac{\alpha_{mn}}{2\pi R}\sqrt{\frac{1}{\rho h}}\sqrt{S+\frac{{\alpha_{mn}}^2 D}{R^2}}.
    \label{eq:f_plate_tension}
\end{equation}
The coefficient $\alpha_{mn}$ corresponds to the $n^{\rm th}$ root of 
\begin{equation}
    \alpha\frac{J_{m+1}(\alpha)}{J_m(\alpha)}+\beta\frac{I_{m+1}(\beta)}{I_m(\beta)}=0,
    \label{eq:alpha_plate}
\end{equation}
where $J_m$ and $I_m$ are the regular and modified Bessel functions of the first kind, respectively, and $\beta=\sqrt{\alpha^2+SR^2/D}$. As $U$ becomes large, e.g., due to a large $S$ or small $h$, the bending term drops out of Eq.~(\ref{eq:eom}), yielding the membrane eigenfrequencies
\begin{equation}
    f_{mn} = \frac{\alpha_{mn}'}{2\pi R}\sqrt{\frac{S}{\rho h}},
    \label{eq:f_membrane}
\end{equation}
where $\alpha_{mn}'$ is the $n^{\rm th}$ root of the Bessel function $J_m$ \cite{weaver1991vibration}.

Figure~\ref{Figure_2}b shows the resonance frequencies of multiple modes of membranes with different $R$ and $h$. Each data point is obtained by averaging measurements from $1 \le N \le 10$ membranes on the same chip with the same linear dimensions. Tables~S1-3 in the Supporting Information file list all measured $f_{mn}$ and $Q_{mn}$. For some membranes, not all modes could be resolved due to low $Q$ values and, in the case of smaller membranes, small displacement amplitudes.

As long as the frequencies of two modes of the same membrane are measured, we can find $E$ and $S$ by performing error minimization in a parametric sweep \cite{ari2020nanomechanical,gress2023multimode}. For each possible combination of $E$ and $S$, we calculate $\alpha_{mn}$ from Eq.~(\ref{eq:alpha_plate}) and the theoretical eigenfrequencies $f_{mn}^{(t)}$ from Eq.~(\ref{eq:f_plate_tension}) for the first four eigenmodes. We then minimize the error between the experimental and theoretical frequencies as described in the Supporting Information file to find the  $E$ and $S$ values.  Figure~\ref{Figure_2}c and d respectively show the $E$ and $S$ found from error minimization. Returning to Figure~\ref{Figure_2}b, the lines show the theoretical $f_{mn}$ based on the mean $E$ and $S$ values for a given $R$ and $h$; the agreement is very good. The values for $E$, $S$, and $U$, as well as the number $N$ of resonators measured for each combination of $R$ and $h$ are listed in Table~\ref{Table_0}. The resonators with $h=65~\rm nm$ have large error bars for $E$ and $S$ due to the small sample size. For the resonators with $h=8~\rm nm$, the rigidity term in Eq.~(\ref{eq:f_plate_tension}) becomes  small compared to the tension term, thus making the tension term dominant and the estimate for $E$ error-prone. In summary, we find the average value $E=11.2\pm8.8~\rm GPa$, which is in good agreement with $E=12.7\pm3.8~\rm GPa$ obtained from nanoindentation measurements \cite{zeng2022irreversible}. We also extracted $E$ and  $S$ using a different approach but arrived at similar values, as  described in the Supporting Information file.

\begin{table}[h]
\caption{Radius $R$, thickness $h$, Young's modulus $E$, tension $S$, and non-dimensional tension parameter $U$ of the resonators in  Figure~\ref{Figure_2}b-d. The data are from  $N$ different resonators on the same chip. \vspace{5pt}} \label{Table_0}
\newcolumntype{Y}{>{\centering\arraybackslash}X}
\begin{tabularx}{0.48\textwidth}{YY|YY|YY|YY|Y}

\multicolumn{2}{c|}{$R\times h$} & \multicolumn{2}{c|}{$E$} & \multicolumn{2}{c|}{$S$} & \multicolumn{2}{c|}{$U$} & $N$\\[2pt]
\multicolumn{2}{c|}{$\left[\rm\upmu m\times nm\right]$} & \multicolumn{2}{c|}{[GPa]} & \multicolumn{2}{c|}{[N/m]} & \multicolumn{2}{c|}{[-]} & [-]\\[3pt]
\hline
\multicolumn{2}{c|}{$~4.25\times8~$} & \multicolumn{2}{c|}{$~25.58\pm6.40~$} & \multicolumn{2}{c|}{$~0.18\pm0.05~$} & \multicolumn{2}{c|}{$~ 3\times 10^3~$} & 6\\
\multicolumn{2}{c|}{$4.25\times15$} & \multicolumn{2}{c|}{$8.42\pm0.28$} & \multicolumn{2}{c|}{$0.39\pm0.01$} & \multicolumn{2}{c|}{$3\times 10^3$} & 5\\
\multicolumn{2}{c|}{$4.25\times35$} & \multicolumn{2}{c|}{$8.44\pm3.55$} & \multicolumn{2}{c|}{$1.37\pm0.02$} & \multicolumn{2}{c|}{$8\times 10^2$} & 10\\
\multicolumn{2}{c|}{$4.25\times65$} & \multicolumn{2}{c|}{$18.07\pm8.06$} & \multicolumn{2}{c|}{$2.26\pm1.09$} & \multicolumn{2}{c|}{$9\times 10^1$} & 3\\
\multicolumn{2}{c|}{$2.75\times35$} & \multicolumn{2}{c|}{$8.80\pm1.40$} & \multicolumn{2}{c|}{$1.41\pm0.04$} & \multicolumn{2}{c|}{$4\times 10^2$} & 3\\
\multicolumn{2}{c|}{$2.75\times65$} & \multicolumn{2}{c|}{$8.20$} & \multicolumn{2}{c|}{$2.20$} & \multicolumn{2}{c|}{$9\times 10^1$} & 1\\
\hline

\end{tabularx}
\end{table}

Figure~\ref{Figure_2}e shows the dissipation constant, $f_{mn} \over Q_{mn}$, for all the measured modes. Even though there is spread in the data from different modes and devices, the dissipation constant seems to increase linearly with frequency, indicating that $Q_{mn}$ is frequency independent. It has been argued \cite{yasumura2000quality} that surface and bulk dissipation respectively result in $Q$ factors that should scale as $Q \propto h$ and $Q \propto \rm constant$. In an effort to observe one of these trends, we plot all $Q_{mn}$ data as a function of $h$ (Figure~\ref{Figure_2}f). The thickest polymer devices appear to have the lowest $Q$ factors, although further experiments are needed for a conclusive statement.

\begin{figure*}
    \includegraphics[width=7in]{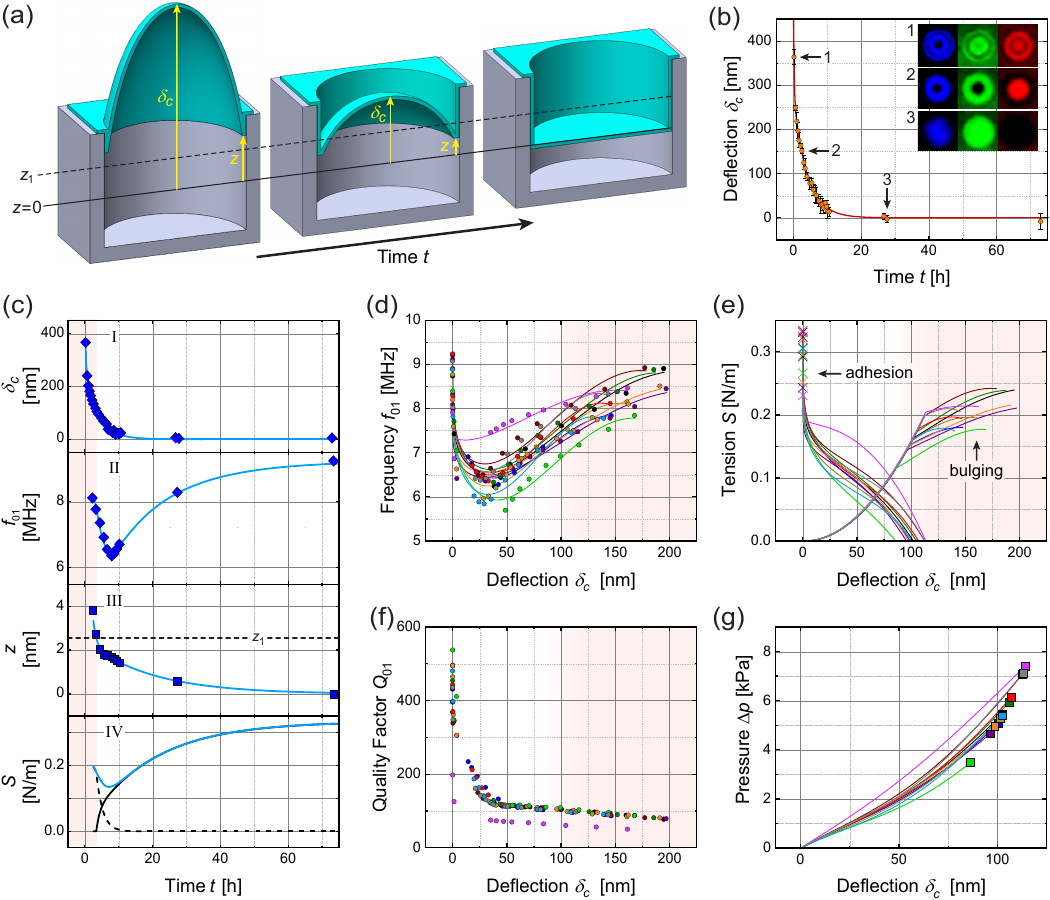}
    \caption{(a) Illustration of a membrane at different stages of bulging. Note the time coordinate of our experiments. When $\Delta p=0$ (right),  the membrane is flat at $z=0$ and stretched. For $\Delta p>0$ (middle), the membrane is adhered to the wall to position $z>0$ and bulges up by $\delta_c-z$. For $z>z_1$ (left), slack is introduced. Note that the illustration is not to scale; in the experiments, $z \ll \delta_c$. (b) Measured center deflection $\delta_c$ for a 35-nm-thick membrane with $R=4.25~\rm\upmu m$ over time. The $\delta_c=0$ line corresponds to the flat membrane at $z=0$. The inset shows interferograms at wavelengths of $440~\rm nm$ (blue), 540~nm (green), and 600~nm (red) for different $\delta_c$ as indicated by the arrows. (c) Time-dependent $\delta_c$, $f_{01}$, $z$, and tension $S$ for a 35-nm thick membrane. The shown values for $z$ are calculated from measurements. The fit for $z$ is an exponential function with two time constants, which is used to calculate the continuous curve in the frequency plot. The dashed line corresponds to the delamination length $z_1$ below which there is no slack. The total tension $S$ (blue line) is comprised of a component due to bulging (black dashed line) and one due to wall adhesion (black solid line). The shading indicates the region of slack in the membrane. (d) Experimental data (symbols) and fits (lines) for $f_{01}$ as a function of $\delta_c$ for twelve 35-nm-thick membranes, including the one in (c). The shaded regions indicate the range where slack exists in the membrane. (e)  Tension components caused by bulging and wall adhesion as a function of $\delta_c$, with the symbols representing the calculated tension at $z=0$. (f) Quality factors $Q_{01}$ as a function of $\delta_c$. (g) Pressure $\Delta p$ in the regime without slack ($z\leq z_1$) as a function of $\delta_c$.}
    \label{Figure_3}
\end{figure*}

To gain more insight into the effect of wall adhesion and delamination on the resonance frequency, we perform a second set of experiments with bulged membranes. First, we describe our model for the resonance of a bulged and partially adhered membrane. The illustration in Figure~\ref{Figure_3}a shows three snapshots of the membrane cross-section along with the relevant coordinates, including the experimental time coordinate. Our mechanical model does not distinguish between inflating and deflating membranes, and we prefer to describe the phenomenon beginning with the final state (backward in time).  At its final flat state ($\Delta p=p_{in}-p_{out}=0$), the membrane is re-adhered to the inner wall of the microwell at $z=0$. The $z_1$ level is selected such that the tension in the membrane  becomes zero if the membrane is flat at $z_1$. As the membrane  separates from the wall starting at $z=0$ toward $z_1$ due to $\Delta p>0$, it maintains some of the initial tension and acquires some additional tension due to the bulging, with $\delta_c$ being the bulge height at the center with respect to $z=0$. When the membrane is at $z_1$, the tension is entirely due to bulging. Further delamination causes a slack being introduced into the membrane, which results in an increased $\delta_c$. The tension in the membrane in these regimes is derived from a spherical cap model in the Supporting Information file and can be summarized as 
\begin{equation}
    S^*(z^*)=\begin{cases}
    z_1^* & \hspace{-0.09\textwidth}\text{for $z^*=0$},\\
    \left(z_1^*-z^*\right)+\frac{2}{3}\left({\delta_c^*-z^*}\right)^2 & \hspace{-0.09\textwidth}\text{for $0<z^*\leq z_1^*$},\\
    \left(z_1^*-z^*\right)+\frac{2}{3}\left(1-z^*+z_1^*\right)\left(\delta_c^*-z^*\right)^2\\  & \hspace{-0.09\textwidth}\text{for $z^*>z_1^*$}.
  \end{cases}
  \label{eq:S_z}
\end{equation}
Here, we present the results in dimensionless form, where $S^*(z^*)=\frac{S(z^*)}{Bh}$, with $B=\frac{E}{1-\nu}$ being the biaxial modulus; the dimensionless length scales are $\delta_c^*=\frac{\delta_c}{R}$, $z^*=\frac{z}{R}$, and $z_1^*=\frac{z_1}{R}$.  Using $S(z)$ in Eqs.~(\ref{eq:alpha_plate}) and (\ref{eq:f_plate_tension}) along with $E$, $\rho$, $\nu$, and linear dimensions  provides the resonance frequency of the bulged-up and delaminated membrane. We reemphasize that the model does not depend on the direction of time.

In our re-adhesion experiments, the membranes are put into a high pressure chamber filled with nitrogen at $p_{ext}=200~\rm kPa$ for several days before being placed into the vacuum chamber ($p_{ext}\approx 0$). We measure the fundamental resonance frequency $f_{01}$ and the quasistatic deflection field simultaneously as gas leaks out of the microwell slowly as a function of time. The deflection field is measured using full-field interferometry \cite{lipiainen2017homodyne} with the details given in the Supporting Information file. In the experiments, the membrane starts in a bulged-up and delaminated state (Figure~\ref{Figure_3}a left). As the gas leaks out of the microwell, the membrane slowly deflates until the final flat state is reached (Figure~\ref{Figure_3}a right). Figure~\ref{Figure_3}b shows the measured deflection $\delta_c$ at the center of a 35-nm-thick membrane over time, which can be fitted with an exponential decay function with two time constants. The asymptote corresponds to a flat membrane ($\delta_c=0$, $z=0$). 

Figure~\ref{Figure_3}c shows measured $\delta_c$ and  $f_{01}$ of a 35-nm-thick membrane as a function of time. Also shown are the calculated position $z$ of the membrane on the wall, and the tension $S$. The analysis is performed as follows. At the last data point ($t\approx 75 ~\rm h$), we assume $z=0$ and determine $S(0)$ from the measured $f_{01}$ using Eqs.~(\ref{eq:f_plate_tension}) and (\ref{eq:alpha_plate}) with $E=12.7~\rm GPa$ and $\nu=0.2$. Then, we find the value for $z_1$ as ${Bh z_1 \over R} = S(0)$. Once $z_1$ is found, we work our way through the rest of the frequency measurements with $\delta_c>0$ and $z>0$. We note that $f_{01}$ depends on $S$ through Eqs. (\ref{eq:alpha_plate}) and (\ref{eq:f_plate_tension}); $S$, in turn, depends on $z$, $\delta_c$ and $z_1$  through Eq.~(\ref{eq:S_z}). Thus, we can determine  the  $z$ value for each $f_{01}$ based on the measured $\delta_c$ and $z_1$ (Figure~\ref{Figure_3}c III).  The data points for $z$ can be fitted with an exponential function with two time constants, which in turn allows us to generate continuous curves for other parameters. With $z$, $\delta_c$ and Eq.~(\ref{eq:S_z}), we also determine the contributions of the tension terms (Figure~\ref{Figure_3}c IV) due to bulging (dotted line) and due to wall adhesion (continuous black line). 

Figure~\ref{Figure_3}d shows the fundamental frequencies $f_{01}$ of twelve 35-nm-thick membranes as a function of $\delta_c$. The continuous curves are found from the model as described above, but plotted as a function $\delta_c$ instead of time. The respective contributions of the two tension terms are shown in Figure~\ref{Figure_3}e. The sudden change in the slope of the bulging plots occurs at $z=z_1$ when slack is introduced to the system. Figure~\ref{Figure_3}f shows the quality factors $Q_{01}$ of the same twelve membranes. As the membranes deflate, $Q_{01}$ increases due to decreased gas damping and increased tension due to adhesion.

Next, we estimate the pressure inside the microwell  by focusing on the regime without slack ($0\leq z\leq z_1$). In this regime, $\Delta p = p_{in}-p_{out} \approx p_{in}$ and can be determined as 
\begin{multline}
    \Delta p= Bh^*(\delta_c^*-z^*)\\\times\left[\frac{8}{3}\left(\delta_c^*-z^*\right)^2+{4}{}\left(z_1^*-z^*\right)+\frac{16}{3(1+\nu)}{h^*}^2\right],
    \label{eq:Delta_p}
\end{multline}
where $h^*={h\over R}$. The derivation for Eq.~(\ref{eq:Delta_p}) can be found in the Supporting Information file. Figure~\ref{Figure_3}g shows $\Delta p$ as a function of $\delta_c$.   

Finally, the energy of adhesion  of the membrane to the substrate can also be estimated, but only at the limit of a flat membrane, i.e., $z\approx 0$. At equilibrium, the total free energy change is
\begin{equation}
    \frac{dW}{dz}+\frac{dF_s}{dz}+\frac{dF_a}{dz}=0,
    \label{eq:equilibrium}
\end{equation}
where all $z$ derivatives are evaluated at $z=0$. Here, $\frac{dW}{dz}$ is the incremental work done by the gas inside the microwell on the membrane; $F_s$ is the strain energy stored in the membrane;  $F_a$ is the free energy of adhesion. Figure~\ref{Figure_3}g shows that  $\frac{dW}{dz}=-\Delta p\frac{dV}{dz} \approx 0$, with $V_b$ being the volume of the bulge, because  $\Delta p \approx 0$ at $z=0$. The strain energy is entirely due to the tension caused by wall adhesion \cite{timoshenko1959theory},
\begin{equation}
    F_s(z) \approx  \pi Bh\left(z_1-z\right)^2,
    \label{eq:F_s}
\end{equation}
and  $\frac{dF_a}{dz}=2\pi R\Gamma$ where $\Gamma$ is the adhesion energy per unit area. Thus, in the limit of $z\approx0$, this free energy model simply yields $\Gamma= Bh\frac{z_1}{R} =S(0)$.  We find the average value of  $\Gamma=0.29\pm0.04~\rm J/m^2$.


The value for $S(0)$ and, therefore, $\Gamma$ is lower for the deflating membranes  (Figure~\ref{Figure_3}e) compared to those that were never bulged by pressurization (Figure~\ref{Figure_2}d).  This observation can  be explained by the history-dependence of  the separation-adhesion process: hysteresis has been observed in separation-adhesion experiments on other 2D materials, where the energy of separation is significantly larger than the energy of adhesion \cite{lloyd2017adhesion}. In our experiments on deflating membranes (Figure~\ref{Figure_3}), we essentially measure the energy of re-adhesion of a membrane that was previously adhered to the substrate and then separated. The data in Figure~\ref{Figure_2} are on membranes that were never separated. It appears that after separation, the membranes do not adhere as strongly, leading to smaller values of tension in their final states. Consistent with this, the deflated membranes in Figure~\ref{Figure_3}f exhibit lower $Q_{01}$ values at $\delta_c=0$ compared with flat membranes of the same radius and thickness that were never pressurized (Figure~\ref{Figure_2}f). Because of higher tension, the  resonance frequencies in Figure~\ref{Figure_2}b are typically higher compared to those in Figure~\ref{Figure_3}d, which also results in higher quality factors due to dissipation dilution~\cite{engelsen2024ultrahigh}.

We assume that equilibrium exists as the membranes re-adhere to the sidewalls of the microwell in the deflation experiments of Figure~\ref{Figure_3}. Here, we take a continuum view of the polymeric material, even though there may be rearrangements in the molecular-scale disks making up the membrane, leading to entropy changes. This is unlike the situation in 2D crystalline materials, such as graphene, and makes the equilibrium in the system more interesting from a fundamental perspective.

In conclusion, we have explored the device possibilities of a novel 2D material, 2DPA-1, by fabricating and measuring  nanomechanical resonators with molecular thicknesses; we have also extracted its material properties and developed complex nanomechanical resonance models with slack and adhesion. Our  $\Gamma$ value for the adhesion of 2DPA-1 on SiO$_2$ is close to those reported for  2D crystalline materials, such as graphene \cite{koenig2011ultrastrong} and MoS$_2$ \cite{lloyd2017adhesion}. The Young's modulus of 2DPA-1, on the other hand, is one to two orders of magnitude smaller \cite{lee2008measurement,castellanos2012elastic}. It remains an open question whether other properties of 2DPA-1 match conventional 1D polymers or crystalline 2D materials.

\section*{Supporting Information}
Additional experimental details on resonance and deflection measurements; information on frequency stability; derivation of the spherical-cap model for delaminating membranes; further information on the determination of material properties; and supporting data tables.

\begin{acknowledgements}
We acknowledge support from the US NSF (Grant Nos. CMMI-2001403, CMMI-1934271, and CMMI-2337507). Polymer syntheses and fabrication of pressurized bulge devices were supported by the Center for Enhanced Nanofluidic Transport–Phase 2 (CENT$^2$), an Energy Frontier Research Center funded by the US Department of Energy, Office of Science, Basic Energy Sciences (Grant No. DE-SC0019112).
\end{acknowledgements}

\onecolumngrid
\begin{center}
\vspace{10pt}
\section*{References}
\vspace{-30pt}
\end{center}
\twocolumngrid

%

\end{document}


\title{Supporting Information for \\ ``Molecularly Thin Polyaramid Nanomechanical Resonators"}

\newcommand{\BU}{Department of Mechanical Engineering, Division of Materials Science and Engineering, and the Photonics Center, Boston University, Boston, Massachusetts 02215, USA}

\author{Hagen Gress}
\affiliation{Department of Mechanical Engineering, Division of Materials Science and Engineering, and the Photonics Center, Boston University, Boston, Massachusetts 02215, USA \looseness=-1}

\author{Cody L. Ritt}
\affiliation{Department of Chemical and Biological Engineering, University of Colorado Boulder, Boulder, Colorado 80309, USA \looseness=-1}

\author{Inal Shomakhov}
\affiliation{Department of Mechanical Engineering, Division of Materials Science and Engineering, and the Photonics Center, Boston University, Boston, Massachusetts 02215, USA \looseness=-1}

\author{Kaan Altmisdort}
\affiliation{Department of Mechanical Engineering, Division of Materials Science and Engineering, and the Photonics Center, Boston University, Boston, Massachusetts 02215, USA \looseness=-1}

\author{Michelle Quien}
\affiliation{Department of Chemical Engineering, Massachusetts Institute of Technology, Cambridge, Massachusetts 02139, USA \looseness=-1}

\author{Zitang Wei}
\affiliation{Department of Chemical Engineering, Massachusetts Institute of Technology, Cambridge, Massachusetts 02139, USA \looseness=-1}

\author{John R. Lawall}
\affiliation{National Institute of Standards and Technology, Gaithersburg, Maryland 20899, USA \looseness=-1}

\author{Narasimha Boddeti}
\affiliation{Washington State University, School of Mechanical and Materials Engineering, Pullman, Washington 99163, USA \looseness=-1}

\author{Michael S. Strano}
\thanks{Email: strano@mit.edu}
\affiliation{Department of Chemical Engineering, Massachusetts Institute of Technology, Cambridge, Massachusetts 02139, USA \looseness=-1}

\author{J. Scott Bunch}
\thanks{Email: bunch@bu.edu}
\affiliation{Department of Mechanical Engineering, Division of Materials Science and Engineering, and the Photonics Center, Boston University, Boston, Massachusetts 02215, USA \looseness=-1}

\author{Kamil L. Ekinci}
\thanks{Email: ekinci@bu.edu}
\affiliation{Department of Mechanical Engineering, Division of Materials Science and Engineering, and the Photonics Center, Boston University, Boston, Massachusetts 02215, USA \looseness=-1}


\maketitle
\widetext
\tableofcontents

\newpage

\setcounter{equation}{0}
\setcounter{figure}{0}
\setcounter{table}{0}
\setcounter{page}{1}
\makeatletter
\renewcommand{\theequation}{S\arabic{equation}}
\renewcommand{\thefigure}{S\arabic{figure}}
\renewcommand{\thetable}{S\arabic{table}}
\renewcommand{\bibnumfmt}[1]{[S#1]}
\renewcommand{\citenumfont}[1]{S#1}

\maketitle

\section{Experimental Details}
\label{section:SI_Experimental_Details}

\subsection{Resonance Measurements}

\begin{figure}[h]
    \includegraphics[width=7in]{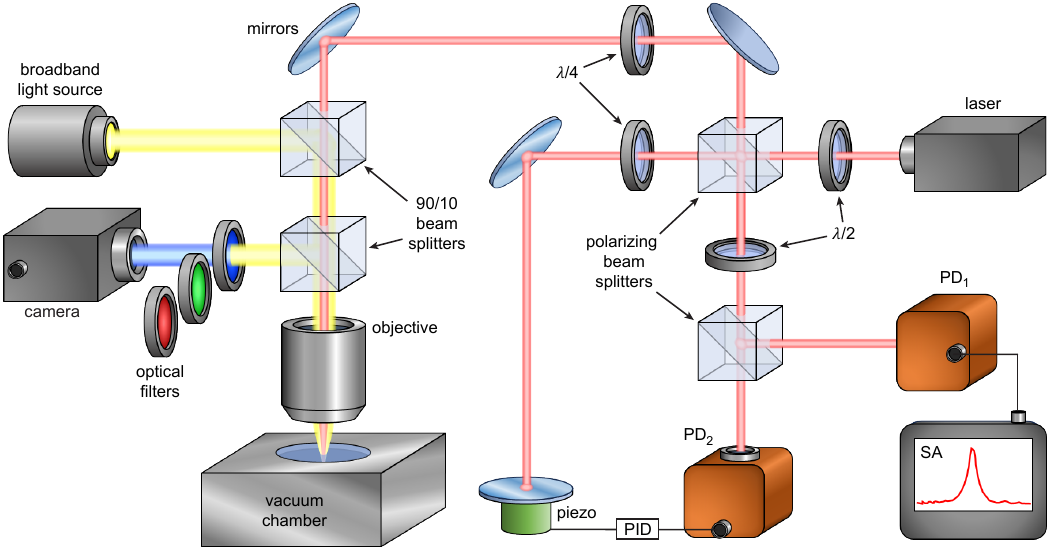}
    \caption{Schematic of the measurement setup. The displacement fluctuations are measured with a path-stabilized homodyne Michelson interferometer. Simultaneously, the static deflection of pressurized membranes is determined via full-field interferometry using a broadband light source, a camera, and three optical filters. $\lambda/2$: half wave plate; $\lambda/4$: quarter wave plate; PD: photodetector; PID: proportional-integral-derivative controller; SA: spectrum analyzer.}    
    \label{Figure_S1}
\end{figure}

We measure the thermal displacement fluctuations of the first four resonant modes with the path-stabilized homodyne Michelson interferometer shown in Figure~\ref{Figure_S1}. We use a stabilized HeNe laser with a wavelength of $\lambda \approx 633~\rm nm$ and a peak power of $3~\rm mW$. The spot size (FWHM) on the sample is about $800~\rm nm$, and the incident power is $\sim600~\rm\upmu W$. We use two photodetectors in our setup. The first detector, PD$_1$, measures the vibrations of the membrane. The second photodetector, PD$_2$, is part of a feedback loop that stabilizes the interferometer at its point of maximum sensitivity, corresponding to an optical path length difference of $\lambda/4$ between the two arms \cite{wagner1990optical}. The bandwidth of the feedback loop is less than $400~\rm Hz$. After taking a measurement at an antinode of the measured mode, we take a second measurement with the same parameters on the polymer film adjacent to the microwell to determine the background noise level in our measurements, e.g., due to the low-frequency laser noise or cable resonances. For both measurements, we average $10^3$ data traces. Based on the assumption that different noise sources are uncorrelated, we subtract our background noise from the noise measured on the membrane.

\subsection{Amplitude Calibration}

To convert the power spectral density (PSD) of displacement fluctuations from the electrical signal recorded with the spectrum analyzer into units of $\rm m^2/Hz$ as shown in Figure~2a in the main paper, we proceed as follows. For each mode $(m, n)$, we determine the modal spring constant as
\begin{equation}
    k_{mn}=\left(2\pi f_{mn}\right)^2m_{mn},
\end{equation}
where $f_{mn}$ is the measured resonance frequency and $m_{mn}$ is the effective modal mass. Assuming that the mode shapes of our resonators can be approximated with those of a circular membrane, we can calculate $m_{mn}$ as \cite{hauer2013general}
\begin{equation}
    \frac{m_{mn}}{\rho\pi R^2h}=\begin{cases}
        \left[J_1\left(\alpha'_{0n}\right)\right]^2 & \text{for $m=0$.}\\
        \frac{K_m^2}{2}\left[J_{m+1}\left(\alpha'_{mn}\right)\right]^2  & \text{for $m>0$.}\\
    \end{cases}
\end{equation}
Here, $\rho$ is the density, $R$ is the radius, and $h$ is the  thickness of   the membrane. $J_m$ are Bessel functions of the first kind and $K_m$ are modified Bessel functions of the second kind. The coefficients $\alpha'_{mn}$ are the $n^{\rm th}$ root of $J_m$.

Once the modal spring constants are known, we  calculate the mean-squared fluctuation amplitude $\left<{W_{mn}}^2\right>$ through the equipartition of energy
\begin{equation}
    \frac{1}{2}k_{mn}\left<{W_{mn}}^2\right>=\frac{1}{2}k_BT,
\end{equation}
where $k_B$ is Boltzmann's constant and $T$ is the temperature.
For the plots shown in Figure~2a, all measurements are multiplied with a coefficient such that for each mode the numerical integral over frequency is equal to $\left<{W_{mn}}^2\right>$.

\subsection{Deflection (Bulge Height) Measurements} \label{subsection:SI_displacement}

\begin{figure}[b]
    \includegraphics[width=7in]{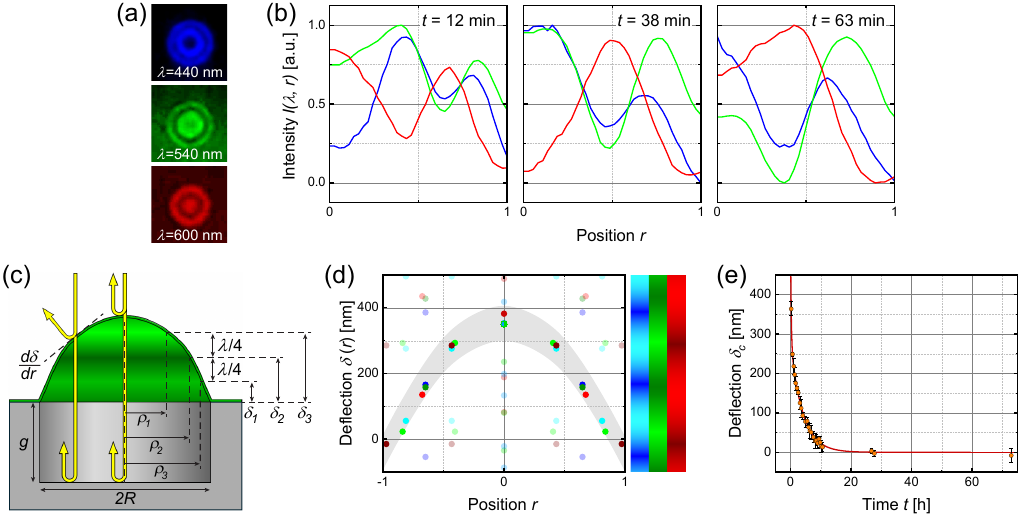}
    \caption{(a) Interferograms of a bulged-up membrane with thickness $h=35~\rm nm$ after the light passes through three different bandpass filters. (b) Radial intensity profiles for each wavelength at three different points in time. The position is normalized by the radius $R$ of the membrane. Local minima correspond to fringes with destructive interference, and local maxima to fringes with constructive interference. (c) Cross-section of a microwell with a bulged-up membrane. The yellow arrows represent how the direction of the reflected light depends on the local slope $d\delta/dr$ of the membrane. The shading of the membrane corresponds to the interference pattern for $\lambda=540~\rm nm$. Each fringe with radius $\rho_i$ corresponds to a deflection $\delta_i$. The vertical distance between to consecutive fringes is always $\lambda/4$. (d) All solutions to Eqs.~(\ref{eq:fringes}) and (\ref{eq:center}). Unphysical values are grayed out. For the fringes, bright colors correspond to maxima, dark colors to minima. A representation of the intensity as a function of deflection is shown on the right. The shaded region represents the extrapolated deflection profile $\delta(r)$ of the bulged-up membrane which is used to identify the correct values $\delta_c$ at the center. (e) Measured center deflection $\delta_c$ of a deflating membrane over time. The data points and error bars are the mean and standard deviation, respectively, of the three values obtained from each wavelength. The red line is an exponential fit.}    
    \label{Figure_S2}
\end{figure}

The semi-transparency of the polymer causes light to be reflected from both the membrane and the bottom of the microwell. We exploit the resulting interference pattern to determine the degree of bulging. Our samples are illuminated with a broadband light source. The reflected light passes through one of three optical bandpass filters with bandwidths $\Delta\lambda\leq16~\rm nm$ and center wavelengths of $\lambda=440$, 540, and 600~nm, before being recorded on a camera (Figure~\ref{Figure_S1}). We calculate the coherence lengths as $L_{\lambda}=\frac{\lambda^2}{\Delta\lambda}$. For our three filters, we find $L_{\lambda}=12.1$, $19.4$, and $25.7~\rm\upmu m$, respectively. The coherence lengths are therefore  larger than the largest difference in optical path length in our experiments, which is $\lesssim3~\rm\upmu m$ for microwells with a depth of $g=960~\rm nm$ and bulges $<500~\rm nm$. Examples of interferograms for each wavelength are shown in Figure~\ref{Figure_S2}a. We use image processing tools to find the average intensities $I(\lambda,r)$ as a function of the radial position $r$ (Figure~\ref{Figure_S2}b). A local maximum corresponds to a constructive (bright) interference fringe and a local minimum to a destructive (dark) one.
The deflection $\delta(r)$ can theoretically be determined at any position from $I(\lambda,r)$ if the intensities at destructive and constructive interference, $I_{min}(\lambda)$ and $I_{max}(\lambda)$, are known \cite{wagner1990optical}. However, values for $I_{min}(\lambda)$ and $I_{max}(\lambda)$ cannot be easily extracted from local extrema because the amount of light that is reflected back along the optical axis depends on the local slope $d\delta/dr$ (Figure~\ref{Figure_S2}c).
We therefore combine two different methodologies to determine $\delta(r)$ at all fringes as well as at the center of the membrane, where $d\delta/dr=0$.

At each interference fringe with radius $\rho$, the deflection is a solution to 
\begin{equation}
\delta(r=\rho)=q\frac{\lambda}{4}-\frac{\lambda}{4\pi}\beta(\lambda) -g,
\label{eq:fringes}
\end{equation}
where the integer $q$ is odd for maxima and even for minima.
The quantity $\beta(\lambda)$ accounts for both the phase shift upon reflection at the membrane and the phase shift acquired by two transmissions through the membrane. For a 35-nm-thick membrane with a refractive index of $n\approx1.57$ \cite{ritt2024molecularly}, we find $\beta(\lambda)=-0.74$, $-0.88$, and $-0.95$ for the three wavelengths, respectively. 

To calculate the deflection $\delta_c$ at the center of the membrane, we observe the intensity $I(\lambda,r=0)$ over time as the membrane gradually deflates (Figure~\ref{Figure_S2}b). As long as the total change in optical path length exceeds one wavelength, we can determine $I_{min}(\lambda)$ and $I_{max}(\lambda)$ as the lowest and highest measured value, respectively. Assuming that the center point is aligned with the optical axis, the center deflection is then a solution to

\begin{equation}
\delta_c=q\frac{\lambda}{4}\pm\frac{\lambda}{4\pi}\cos^{-1}\left(1-2\frac{I(\lambda, r=0)-I_{min}(\lambda)}{I_{max}(\lambda)-I_{min}(\lambda)}\right) -\frac{\lambda}{4\pi}\beta(\lambda) -g,
\label{eq:center}
\end{equation}
where $q$ is an even integer. We next identify the solutions to Eqs.~(\ref{eq:fringes}) and (\ref{eq:center}) which match the parabolic deflection profile $\delta(r)$ that we expect based on theory and AFM measurements. To that end, we analyze the interference pattern for each wavelength as follows. Among the solutions to Eq.~(\ref{eq:fringes}) for the outermost fringe, we select the value for $q$ that corresponds to the smallest deflection. Working our way towards the center, we increase $q$ by 1 for each subsequent fringe. At the center, we assume that the deflection is within the range of $\delta(r=\rho^*)<\delta_c<\delta(r=\rho^*)+\frac{\lambda}{4}$, where $\rho^*$ is the radius of the smallest observable fringe. We extrapolate $\delta(r)$ (shaded region in Figure~\ref{Figure_S2}d) to confirm the validity of our calculations. Our fringe deflections $\delta(r=\rho)$ therefore allow us to identify the correct center deflections $\delta_c$. The reported bulge height is the average of the $\delta_c$ values obtained for each wavelength, with a typical standard deviation of $\sim10~\rm nm$. 

Figure~\ref{Figure_S2}e shows the center deflection $\delta_c$ of a deflating membrane over time. The data is fit with an exponential function with two time constants. Errors in the optical path length difference due to wall adhesion (main text Figure 1d) or deviations from the nominal microwell depth $g$ can result in an offset of all data points in Figure~\ref{Figure_S2}d along the $y$-axis. We correct for these errors by setting the asymptote of the exponential fit, i.e., the flat membrane, as our reference plane ($\delta_c=0$ at $t\rightarrow\infty$).

\begin{figure}[h]
    \includegraphics[width=7in]{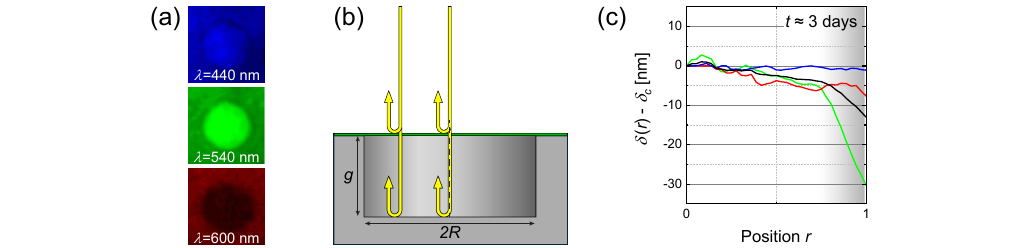}
    \caption{(a) Interferograms of a presumed flat membrane with thickness $h=35~\rm nm$ after the light passes through three different bandpass filters. (b) Cross-section of a microwell with a flat membrane. The yellow arrows represent how, unlike in a bulged membrane, the amount of light reflected along the optical axis is independent of the radial position $r$. (c) Deflection profile of a nominally flat membrane relative to its center. The black line represents the average over the three color curves. The gray region indicates where proximity to the wall of the microwell renders the intensity measurement unreliable.}    
    \label{Figure_S2_2}
\end{figure}

To verify that the membranes are fully deflated at the end of the experiment, we turn to the inteferograms corresponding to the last data point (Figure~\ref{Figure_S2_2}a). We assume that $d\delta/dr\approx0$ across the whole membrane. Therefore, the amount of incident light being reflected back along the optical axis should become independent of position $r$ (Figure~\ref{Figure_S2_2}b). Any deviation from the center deflection can then be estimated as
\begin{equation}
\delta(r)-\delta_c=\pm\frac{\lambda}{4\pi}\left[\cos^{-1}\left(1-2\frac{I(\lambda, r)-I_{min}(\lambda)}{I_{max}(\lambda)-I_{min}(\lambda)}\right)-\cos^{-1}\left(1-2\frac{I(\lambda, r=0)-I_{min}(\lambda)}{I_{max}(\lambda)-I_{min}(\lambda)}\right)\right].
\label{eq:flat}
\end{equation}
Due to the fact that $\delta_c$ is monotonically decreasing, the sign in Eq.~(\ref{eq:flat}) is positive when $I_{max}$ was the last extremum observed during deflation and negative when $I_{min}$ was the last extremum. Figure~\ref{Figure_S2_2}c shows a deflection measurement of a  membrane that is flat. Because the wall of the microwell interferes with the intensity measurements near the edge of the membrane, we focus on the data for $r\lesssim0.75$ and find that our current setup can determine flatness to within approximately $10~\rm nm$.

\newpage
\section{Frequency Stability}

To assess the stability of our resonators, we measured the resonance frequencies $f_{01}$ and quality factors $Q_{01}$ of three 35-nm-thick membranes over three days in air and fourteen days in vacuum. In air, $f_{01}$ and $Q_{01}$ were stable to within $3\%$ and $5\%$, respectively. After five days in vacuum, the resonance frequencies stabilized to within $10\%$, whereas the quality factors continuously drifted upward, increasing by $50-60\%$.

\newpage
\section{Spherical Cap Model}

\subsection{Tension per Unit Length for a Slacked Membrane}
\begin{figure}[h]
    \includegraphics[width=7in]{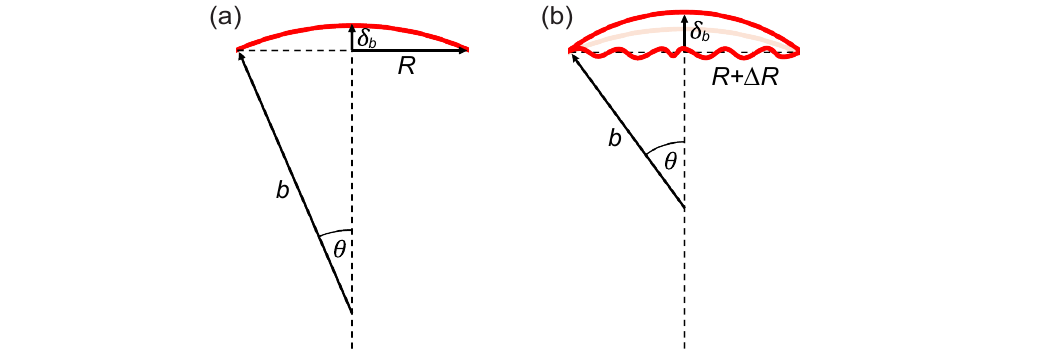}
    \caption{(a) Spherical cap, where the microwell radius is $R$, the radius of  curvature of the bulge is $b$, and the  height of the bulge at the center is $\delta_b$. We assume  that $b\gg R$ and  $b \gg \delta_b$. (b) For a slacked membrane, the effective radius exceeds the radius of the microwell by a small length $\Delta R$. This leads to a larger value for $\delta_b$ when the membrane is pressurized.}    
    \label{Figure_S3}
\end{figure}

Figure~\ref{Figure_S3}a shows the  spherical cap geometry for a bulging membrane  on a cylindrical microwell of radius $R$ without any slack or pre-tension. We first derive the relationship between the radius of curvature $b$ and the bulge height $\delta_b$ at the center  through the Pythagorean theorem for the limit $b\gg R$ and $b\gg \delta_b$:
\begin{equation}
    b^2=R^2+(b-\delta_b)^2\approx R^2+b^2-2\delta_b b.
\end{equation}
Thus,
\begin{equation}
    b \approx \frac{R^2}{2\delta_b}.
    \label{eq:b_R}
\end{equation}
The biaxial strain $\varepsilon$ for this case is 
\begin{equation}
    \varepsilon\approx\frac{b\theta-R}{R}.
\end{equation}
We express the angle $\theta$ as a Taylor series expansion as
\begin{equation}
    \theta=\arcsin{\left(\frac{R}{b}\right)}=\frac{R}{b}+\frac{1}{6}\left(\frac{R}{b}\right)^3+...
\end{equation}
The strain can then be found as
\begin{equation}
    \varepsilon\approx\frac{b \left( \frac{R}{b}+\frac{1}{6}\frac{R^3}{b^3}\right)-R}{R}\approx\frac{R^2}{6b^2}.
\end{equation}
Substituting for $b$ from Eq.~(\ref{eq:b_R}) above, 
\begin{equation}
    \varepsilon \approx\frac{R^2}{6\left(\frac{R^2}{2\delta_b}\right)^2}\approx\frac{2}{3}\left(\frac{\delta_b}{R}\right)^2.
    \label{eq:epsilon_bulge}
\end{equation}

For a membrane with slack  shown in Figure~\ref{Figure_S3}b, the strain is dependent on the effective radius of the membrane, $R+\Delta R$,  and can be approximated as
\begin{equation}
\begin{split}
    \varepsilon & \approx\frac{b\theta-(R+\Delta R)}{R+\Delta R}\\
    &\approx\frac{b\left(\frac{R}{b}+\frac{1}{6}\frac{R^3}{b^3}\right)-(R+\Delta R)}{R+\Delta R}\\
    &\approx\frac{R^2}{6b^2\left(1+\frac{\Delta R}{R}\right)}-\frac{\Delta R}{R\left(1+\frac{\Delta R}{R}\right)}\\
    & \approx\frac{R^2}{6b^2}-\frac{R\Delta R}{6b^2}-\frac{\Delta R}{R}.\\
\end{split}
\end{equation}
Substituting for $b$ again,
\begin{equation}
\begin{split}
    \varepsilon &\approx\frac{R^2}{6\left(\frac{R^2}{2\delta_b}\right)^2}-\frac{R\Delta R}{6\left(\frac{R^2}{2\delta_b}\right)^2}-\frac{\Delta R}{R}\\
    &\approx\frac{2}{3}\left(1-\frac{\Delta R}{R}\right)\left(\frac{\delta_b}{R}\right)^2-\frac{\Delta R}{R}.
    \label{eq:epsilon_bulge_slack}
    \end{split}
\end{equation}
The tension per unit length $S$ is related to the strain by
\begin{equation}
    S=\frac{Eh}{1-v}\varepsilon.
\end{equation}

\subsection{Application to the Delaminating Membrane}
\begin{figure}[t]
    \includegraphics[width=7.0in]{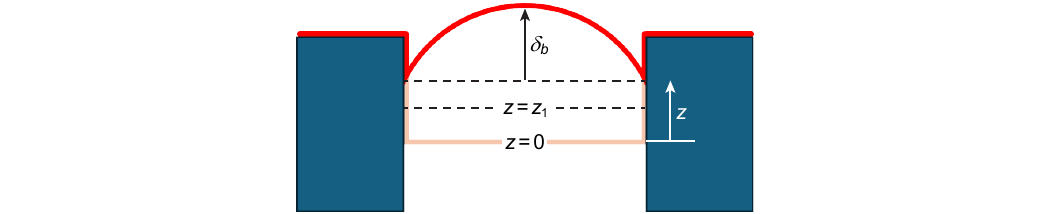}
    \caption{Schematic of a pressurized membrane in a microwell. At the vertical position $z=0$, the membrane is completely flat ($\Delta p=0$) and under tension due to wall adhesion. When the membranes delaminates a distance $z_1$ due to an applied pressure ($\Delta p>0$), neither tension due to wall adhesion nor slack exist. For $z>z_1$, slack leads to an increased bulge height $\delta_b$.}    
    \label{Figure_S4}
\end{figure}

In this system, the  slack depends on the degree of delamination, as illustrated in Figure \ref{Figure_S4}. Thus, we define a coordinate system $z$, where $z=0$ corresponds to the vertical position of the flat membrane ($\Delta p=0$). The strain in the flat membrane is entirely caused by wall adhesion and can be expressed as $\varepsilon=z_1/R$, where $z_1$ is the coordinate along the wall such that the tension in the membrane becomes zero if the membrane is flat at that level.  When pressure is applied ($\Delta p>0$), the membrane delaminates a distance $z$. The bulge height is then $\delta_b=\delta_c-z$, where $\delta_c$ is the measured center deflection with respect to the flat state. The delamination reduces the strain component due to adhesion to $\varepsilon=(z_1-z)/R$ while a second tension component due to bulging, as found in Eq.~(\ref{eq:epsilon_bulge}), is introduced. At the position $z_1$, there is no more tension due to wall adhesion. Any further delamination introduces slack and the effective radius of the membrane is increased by $\Delta R=z-z_1$. The strain for $z>z_1$, where a net slack of $\Delta R=z-z_1$ is introduced into the membrane,  is determined from Eq.~(\ref{eq:epsilon_bulge_slack}).

Taking these into account and using the $\varepsilon$ expressions found above, the tension can be calculated for different regimes of $z$ as follows:

\begin{equation}
  S(z)=\begin{cases}
    \frac{Eh}{1-\nu}\left(\frac{z_1}{R}\right) & \text{for $z=0$}.\\
    \frac{Eh}{1-\nu}\left[\left(\frac{z_1-z}{R}\right)+\frac{2}{3}\left({\delta_c-z}\over R\right)^2 \right] & \text{for $0<z\leq z_1$}.\\
    \frac{Eh}{1-\nu}\left[\left({z_1-z}\over R \right)+\frac{2}{3}\left(1-\frac{z-z_1}{R}\right)\left({\delta_c-z} \over R\right)^2\right] & \text{for $z>z_1$}.
  \end{cases}
  \label{eq:S_z}
\end{equation}

We prefer to rewrite the strain and all the length scales in dimensionless units. The dimensionless length scales that we select are $z^*={z \over R}$, $z_1^*={z_1 \over R}$, and $\delta_c^*={\delta_c \over R}$.  
\begin{equation}
  {S(z^*) \over \left(\frac{Eh}{1-\nu} \right)} =\begin{cases}
    z_1^* & \text{for $z^*=0$}.\\
    \left(z_1^*-z^*\right)+\frac{2}{3}\left({\delta_c^*-z^*}\right)^2 & \text{for $0<z^*\leq z_1^*$}.\\
  z_1^*-z^*+  \frac{2}{3}\left(1-z^*+
    z_1^*\right)\left(\delta_c^*-z^*\right)^2 & \text{for $z^*>z_1^*$}.
  \end{cases}
\end{equation}
In the experiments, the dimensionless length scales typically vary as follows: $0 \lesssim z^* \lesssim 10^{-3}$; $z_1^*\approx10^{-3}$; and $0 \lesssim \delta_c^* \lesssim 10^{-1}$. 
\subsection{Pressure inside the Spherical Cap with No Slack}

Assuming biaxial strain, the pressure can be found as follows.
The stress $\sigma$ in a spherical cap with radius $b$ and thickness $h$ can be related to the pressure $\Delta p= p_{in}-p_{out}$ by balancing forces, leading to
\begin{equation}
    2\pi Rh\sigma \sin{\theta}=\pi R^2\Delta p.
\end{equation}
From $\sin{\theta}=\frac{R}{b}$ and Eq.~(\ref{eq:b_R}), we  find
\begin{equation}
    \Delta p=\frac{2h}{b}\sigma=\frac{4h(\delta_c-z)}{R^2}\sigma.
    \label{eq:p_sigma}
\end{equation}
The stress is related to the strain by
\begin{equation}
    \sigma=\frac{E}{1-\nu}\varepsilon=\frac{S}{h}
    \label{eq:sigma_S}
\end{equation}
and therefore has two components: one due to wall adhesion, the other one due to bulging. For $z\leq z_1$, we can combine Eqs.~(\ref{eq:S_z}), (\ref{eq:p_sigma}), and (\ref{eq:sigma_S}) to determine the pressure as
\begin{equation}
    \Delta p=\frac{8Eh}{3(1-\nu)R}\left(\frac{\delta_c-z}{R}\right)^3+\frac{4Eh}{(1-\nu)R}\left(\frac{z_1-z}{R}\right)\left({\delta_c-z} \over R\right).
    \label{eq:p_delta}
\end{equation}
We add a term to Eq.~(\ref{eq:p_delta}) to account for the bending rigidity \cite{poilane2000analysis}, leading to
\begin{equation}
    \Delta p=\frac{8Eh}{3(1-\nu)R}\left(\frac{\delta_c-z}{R}\right)^3+\frac{4Eh}{(1-\nu)R}\left(\frac{z_1-z}{R}\right)\left(\frac{\delta_c-z}{R}\right)+\frac{16Eh^3}{3(1-\nu^2)R^3}\left(\frac{\delta_c-z}{R}\right).
\end{equation}
Using the dimensionless length scales as above and defining $h^*={h \over R}$, we write
\begin{equation}
    \Delta p=\frac{8E}{3(1-\nu)}h^*\left(\delta_c^*-z^*\right)^3+\frac{4E}{(1-\nu)} h^*\left(z_1^*-z^*\right)(\delta_c^*-z^*)+\frac{16E}{3(1-\nu^2)} {h^*}^3(\delta_c^*-z^*).
\end{equation}
This simplifies to
\begin{equation}
    \Delta p=\frac{Eh^*(\delta_c^*-z^*)}{1-\nu}\left[\frac{8}{3}\left(\delta_c^*-z^*\right)^2+{4}{}\left(z_1^*-z^*\right)+\frac{16}{3(1+\nu)}{h^*}^2\right].
\end{equation}
The typical values for the dimensionless length scales for $z\leq z_1$ are as follows: $h^* \approx 10^{-2}$; $0 \lesssim z^* \lesssim 10^{-3}$; $z_1^*\approx10^{-3}$; $0 \lesssim \delta_c^* \lesssim 10^{-2}$. In the limit of $z\to z_1$ ($z^*\to z_1^*$), we find that $(\delta_c^*-z^*)\sim h^*$, so both the first and third terms inside the brackets need to be taken into account.

\subsection{Strain Energy}

In general, the strain energy for the delaminating membrane with \textit{no} slack should be  a complicated function of $z$ and $\delta_c$, given that the membrane is stretched both due to  bulging and  wall adhesion.  Here, we will find approximate expressions for the strain energy \textit{only} at the limit of a flat membrane.

For a biaxially strained nearly flat membrane around  $z=0$, the strain energy can be approximated as
\begin{equation}
    F_s(z) \approx S(z)\varepsilon(z)\pi R^2.
\end{equation}
We do not consider any deformation or stored strain energy in the portion of the membrane that is adhered to the sidewall. We substitute for $S(z)$ and $\varepsilon(z)$ and find
\begin{equation}
    F_s(z) \approx  \frac{Eh}{1-\nu}\left(\frac{z_1-z}{R}\right)^2\pi R^2.
\end{equation}
Here, we completely neglected the contribution of the bulging to the energy. We justify this assumption by noting that, as $z\to 0$, $\delta_c \to 0$ and the quadratic contribution of the second term in Eq.~(\ref{eq:S_z}) to strain becomes even smaller.

\newpage
\section{Determining $E$, $S$ and $\nu$ }

\subsection{Searching for Both $E$ and $S$ with Fixed $\nu=0.2$}
To find the Young's modulus $E$ and the tension $S$, we perform a parametric sweep. For each possible combination of $E$ and $S$, we first calculate the coefficients $\alpha$ and $\beta$ from
\begin{equation}
    \alpha\frac{J_{m+1}(\alpha)}{J_m(\alpha)}+\beta\frac{I_{m+1}(\beta)}{I_m(\beta)}=0,
    \label{eq:alpha_plate}
\end{equation}
where $J_m$ and $I_m$ are the regular and modified Bessel functions of the first kind, respectively, and $\beta=\sqrt{\alpha^2+SR^2/D}$. We then calculate the theoretical eigenfrequencies
\begin{equation}
    f_{mn}^{(t)} = \frac{\alpha_{mn}}{2\pi R}\sqrt{\frac{1}{\rho h}}\sqrt{S+\frac{{\alpha_{mn}}^2 D}{R^2}}.
    \label{eq:f_plate_tension}
\end{equation}
for the first four eigenmodes. The density $\rho=1140~\rm kg/m^3$ \cite{ritt2024molecularly} as well as the dimensions $R$ and $h$ are known; we take $\nu=0.2$, assuming a Poisson's ratio similar to that of graphene \cite{politano2015probing}. For each measured eigenmode, we calculate the magnitude of the error between the measured and theoretical frequencies as
\begin{equation}
    \epsilon_{mn}= \sqrt{\frac{\left(f_{mn}-f_{mn}^{(t)}\right)^2}{{f_{mn}}^2}}.
    \label{eq:error}
\end{equation}
We then find the total (squared) error by summing ${\epsilon_{mn}}^2$ over all the measured modes. The two-dimensional sweep of $E$ and $S$ reveals the values that minimize the total error. We run the error minimization algorithm for each measured membrane separately, because the value of $S$ is expected to be different from sample to sample.

\subsection{Searching for $S$ with Fixed $E=12.7~\rm GPa$ and $\nu=0.2$} \label{subsection:E}

\begin{figure}[h]
    \includegraphics[width=7in]{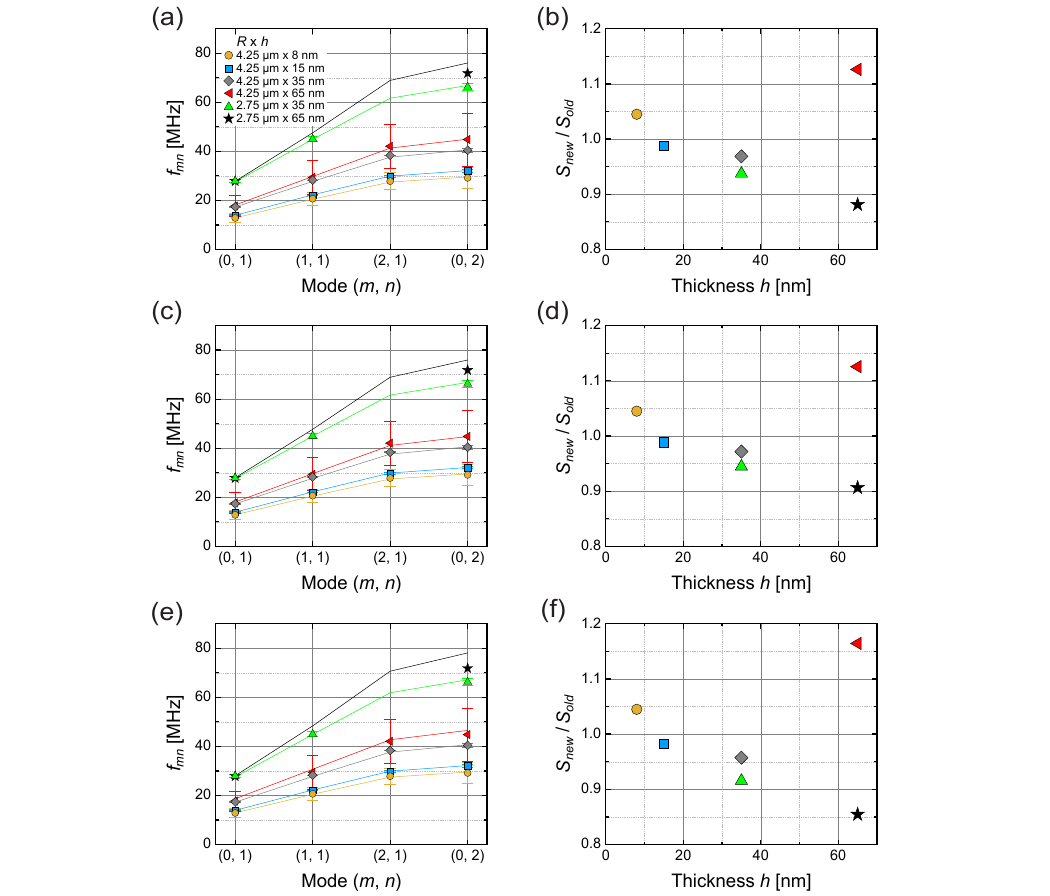}
    \caption{(a) Resonance frequencies $f_{mn}$ for membranes with indicated $R$ and $h$, assuming a Young's modulus $E=12.7~\rm GPa$ and a Poisson's ratio $\nu=0.2$. (b) Ratio of the tension values $S_{new}$ when $E$ is fixed to the tension values $S_{old}$ reported in the main paper when $E$ is also swept. (c) Resonance frequencies $f_{mn}$ and (d) tension ratio $S_{new}/S_{old}$ when the Poisson's ratio is set to $\nu=0.1$. (e) Resonance frequencies $f_{mn}$ and (f) tension ratio $S_{new}/S_{old}$ when the Poisson's ratio is set to $\nu=0.4$.}    
    \label{Figure_S5}
\end{figure}

Figure~2 in the main text shows a large scatter for the Young's modulus $E$. We therefore investigate the effect of $E$ on the calculated tension $S$. We take $E=12.7$ \cite{zeng2022irreversible}, sweep the value for the tension $S$, and calculate the corresponding theoretical eigenfrequencies $f_{mn}^{(t)}$ based on Eqs.~(\ref{eq:alpha_plate}) and (\ref{eq:f_plate_tension}).
We calculate the error between the experimental and theoretical eigenfrequencies for each value of $S$ with Eq.~(\ref{eq:error}).
We find $S$ by minimizing the total error. Figure~\ref{Figure_S5}a shows the resulting calculated and measured frequencies. Figure~\ref{Figure_S5}b shows the ratio between the new tension values $S_{new}$ when $E$ is fixed and the old values  $S_{old}$ reported in the main text, when $E$ is also swept. Fixing $E=12.7~\rm GPa$ changes the calculated tension by $<15\%$.
For our thinnest membranes, the bending stiffness becomes negligible. Instead of using Eq.~(\ref{eq:f_plate_tension}), we therefore compare the experimental frequencies of the 8-nm-thick membranes to those of an ideal membrane
\begin{equation}
    f_{mn}^{(t)} = \frac{\alpha_{mn}'}{2\pi R}\sqrt{\frac{S}{\rho h}},
    \label{eq:f_membrane}
\end{equation}
where $\alpha_{mn}'$ is the $n^{\rm th}$ root of the Bessel function $J_m$ \cite{weaver1991vibration}. The calculated tension differs from the value reported in the main paper by $<5\%$.

\subsection{Effect of Changing $\nu$}

In the main text, we assume that 2DPA-1 has a Poisson ratio similar to graphene and set $\nu=0.2$. Here, we investigate how much the calculated $S$ depends on the assumed value for $\nu$. We therefore repeat the steps as outlined in Section~\ref{subsection:E} at a fixed $E=12.7~\rm GPa$ with $\nu=0.1$ (Figure~\ref{Figure_S5}c and \ref{Figure_S5}d) and $\nu=0.4$ (Figure~\ref{Figure_S5}e and \ref{Figure_S5}f). The results for $\nu=0.1$ are almost identical to those for $\nu=0.2$. For $\nu=0.4$, the 65-nm-thick membranes are affected the most, and $S$ changes up to $18\%$ compared to the value reported in the main text.
\clearpage
\newpage

\section{Data Tables}

\begin{table}[h]
\begin{minipage}{\textwidth}
\caption{Frequencies $f_{mn}$ and quality factors $Q_{mn}$ for the larger resonators with radius $R=4.25~\rm \upmu m$ that were used to calculate the Young's modulus $E$ and the tension $S$. \vspace{5pt}} \label{Table_1}
\renewcommand{\arraystretch}{1}
 \newcolumntype{Y}{>{\centering\arraybackslash}X}
\begin{tabularx}{\textwidth}{Y|Y|YY|YY|YY|YY|Y|Y}

\multicolumn{2}{c|}{} & \multicolumn{2}{c|}{Mode (0, 1)} & \multicolumn{2}{c|}{Mode (1, 1)} & \multicolumn{2}{c|}{Mode (2, 1)} & \multicolumn{2}{c|}{Mode (0, 2)} & \multicolumn{2}{c}{}\\[3pt]
\hline
$h$ & Resonator & $f_{01}$ & $Q_{01}$ & $f_{11}$ & $Q_{11}$ & $f_{21}$ & $Q_{21}$ & $f_{02}$ & $Q_{02}$ & $E$ & $S$\\[2pt]
[nm] & & [MHz] & [-] & [MHz] & [-] & [MHz] & [-] & [MHz] & [-] & [GPa] & [N/m]\\[3pt]
\hline
8 & 1 & 10.249 & 914 & 16.396 & 917 & 22.593 & 849 & 23.042 & 640 & 16.5 & 0.11\\
&  &  &  & 16.533 & 749 & 22.714 & 785 &  &  & \\
& 2 & 14.344 & 712 & 22.839 & 826 & 30.700 & 904 & 32.632 & 884 & 31.0 & 0.22\\
& &  &  & 22.867 & 867 & 30.896 & 924 &  &  & \\
& 3 & 13.396 & 364 & 21.308 & 485 & 28.549 & 535 & 30.565 & 501 & 27.0 & 0.19\\
& &  &  & 21.349 & 516 & 28.764 & 455 &  &  &  & \\
& 4 & 13.118 & 887 & 21.842 & 1002 & 28.677 & 725 & 29.977 & 958 & 27.9 & 0.19\\
&  &  &  & 22.111 & 1009 & 29.786 & 1026 &  &  &  & \\
& 5 & 11.202 & 1302 & 17.900 & 1107 & 24.319 & 860 & 25.236 & 783 & 19.0 & 0.14\\
&  &  &  & 17.922 & 1076 & 24.433 & 902 &  &  &  & \\
& 6 & 14.577 & 1600 & 23.221 & 1555 & 31.052 & 1423 & 33.452 & 1582 & 32.1 & 0.23\\
&  &  &  & 23.267 & 1585 & 31.324 & 1591 &  &  &  & \\
15 & 1 & 13.609 & 969 & 21.660 & 877 & 28.926 & 798 & 31.276 & 776 & 8.0 & 0.37\\
&  &  &  & 21.656 & 972 & 29.167 & 935 &  &  &  & \\
& 2 & 14.225 & 658 & 22.468 & 870 & 30.120 & 737 & 32.609 & 665 & 8.7 & 0.41\\
&  &  &  & 22.743 & 610 & 30.440 & 705 &  &  &  & \\
& 3 & 14.081 & 1004 & 22.383 & 1113 & 29.915 & 988 & 32.356 & 898 & 8.5 & 0.40\\
&  &  &  & 22.427 & 1098 & 30.172 & 1000 &  &  &  & \\
& 4 & 13.868 & 1170 & 22.039 & 1230 & 29.504 & 1197 & 31.874 & 1086 & 8.3 & 0.39\\
&  &  &  & 22.118 & 1253 & 29.715 & 1091 &  &  &  & \\
& 5 & 14.085 & 1002 & 22.382 & 1103 & 29.955 & 1085 & 32.374 & 981 & 8.6 & 0.40\\
&  &  &  & 22.456 & 1151 & 30.175 & 933 &  &  &  & \\
35 & 1 & 17.601 & 840 & 28.199 & 666 & 38.203 & 542 & 41.324 & 484 & 13.2 & 1.37\\
&  &  &  & 28.289 & 660 & 38.457 & 504 &  &  &  & \\
& 2 & 17.440 & 786 & - & - & - & - & 40.907 & 488 & 11.9 & 1.36\\
&  &  &  & - & - & - & - &  &  &  & \\
& 3 & 17.386 & 737 & - & - & - & - & 40.758 & 427 & 11.4 & 1.35\\
&  &  &  & - & - & - & - &  &  &  & \\
& 4 & 17.150 & 976 & - & - & - & - & 39.877 & 669 & 6.9 & 1.35\\
&  &  &  & - & - & - & - &  &  &  & \\
& 5 & 17.119 & 845 & - & - & - & - & 39.704 & 608 & 5.6 & 1.35\\
&  &  &  & - & - & - & - &  &  &  & \\
& 6 & 17.005 & 846 & - & - & - & - & 39.416 & 600 & 5.2 & 1.34\\
&  &  &  & - & - & - & - &  &  &  & \\
& 7 & 17.376 & 1493 & - & - & - & - & 40.255 & 965 & 5.3 & 1.40\\
&  &  &  & - & - & - & - &  &  &  & \\
& 8 & 17.307 & 1224 & - & - & - & - & 40.100 & 860 & 5.3 & 1.39\\
&  &  &  & - & - & - & - &  &  &  & \\
& 9 & 17.229 & 1043 & - & - & - & - & 40.002 & 731 & 6.2 & 1.37\\
&  &  &  & - & - & - & - &  &  &  & \\
& 10 & 17.751 & 987 & - & - & - & - & 41.692 & 552 & 13.4 & 1.40\\
&  &  &  & - & - & - & - &  &  &  & \\
65 & 1 & 22.426 & 347 & 37.074 & 249 & 52.095 & 194 & 57.173 & 170 & 27.1 & 3.52\\
&  &  &  & 37.437 & 237 & 52.517 & 218 &  &  &  & \\
& 2 & 15.010 & 244 & 24.880 & 165 & 35.965 & 138 & 37.310 & 125 & 11.6 & 1.60\\
&  &  &  & - & - & - & - &  &  &  & \\
& 3 & 15.659 & 177 & 26.079 & 107 & 37.936 & 107 & 39.940 & 100 & 15.5 & 1.65\\
&  &  &  & - & - & - & - &  &  &  & \\
\hline

\end{tabularx}
\end{minipage}
\end{table}

\begin{table}[h]
\begin{minipage}{\textwidth}
\caption{Frequencies $f_{mn}$ and quality factors $Q_{mn}$ for the smaller resonators with radius $R=2.75~\rm \upmu m$ that were used to calculate the Young's modulus $E$ and the tension $S$. For mode (0, 2), the quality factors could not be determined due to a low signal-to-noise ratio. \vspace{5pt}} \label{Table_2}
\renewcommand{\arraystretch}{1}
 \newcolumntype{Y}{>{\centering\arraybackslash}X}
\begin{tabularx}{\textwidth}{Y|Y|YY|YY|YY|YY|Y|Y}

\multicolumn{2}{c|}{} & \multicolumn{2}{c|}{Mode (0, 1)} & \multicolumn{2}{c|}{Mode (1, 1)} & \multicolumn{2}{c|}{Mode (2, 1)} & \multicolumn{2}{c|}{Mode (0, 2)} & \multicolumn{2}{c}{}\\[3pt]
\hline
$h$ & Resonator & $f_{01}$ & $Q_{01}$ & $f_{11}$ & $Q_{11}$ & $f_{21}$ & $Q_{21}$ & $f_{02}$ & $Q_{02}$ & $E$ & $S$\\[2pt]
[nm] & & [MHz] & [-] & [MHz] & [-] & [MHz] & [-] & [MHz] & [-] & [GPa] & [N/m]\\[3pt]
\hline
35 & 1 & 27.623 & 579 & 44.168 & 387 & - & - & 65.111 & - & 6.6 & 1.40\\
&  &  &  & - & - & - & - &  &  &  & \\
& 2 & 28.557 & 460 & 45.783 & 312 & - & - & 67.995 & - & 9.4 & 1.46\\
&  &  &  & - & - & - & - &  &  &  & \\
& 3 & 27.582 & 389 & - & - & - & - & 65.508 & - & 8.0 & 1.38\\
&  &  &  & - & - & - & - &  &  &  & \\
65 & 1 & 27.825 & 271 &  &  &  &  & 71.846 &  & 8.2 & 2.20\\
&  &  &  & - & - & - & - &  &  &  & \\
\hline

\end{tabularx}
\end{minipage}
\end{table}

\begin{table}[h]
\begin{minipage}{\textwidth}
\caption{Frequencies $f_{01}$ and quality factors $Q_{01}$ for additional resonators for which only the fundamental mode was measured. \vspace{5pt}} \label{Table_3}
\renewcommand{\arraystretch}{1}
 \newcolumntype{Y}{>{\centering\arraybackslash}X}
\begin{tabularx}{\textwidth}{Y|Y|Y|YY|YYY}

\multicolumn{3}{c|}{} & \multicolumn{2}{c|}{Mode (0, 1)} & \multicolumn{3}{c}{}\\[3pt]
\hline
$h$ & $R$ & Resonator & $f_{01}$ & $Q_{01}$ & \multicolumn{3}{c}{comment} \\[2pt]
[nm] & $\upmu m$ & & [MHz] & [-] & & & \\[3pt]
\hline

35 & 4.25 & 1 & 8.735 & 396 & \multicolumn{3}{c}{Figures~3d and 3f at $\delta_c=0~\rm nm$} \\
& & 2 & 9.114 & 431 & \multicolumn{3}{c}{Figures~3d and 3f at $\delta_c=0~\rm nm$} \\
& & 3 & 9.207 & 433 & \multicolumn{3}{c}{Figures~3d and 3f at $\delta_c=0~\rm nm$} \\
& & 4 & 8.904 & 454 & \multicolumn{3}{c}{Figures~3d and 3f at $\delta_c=0~\rm nm$} \\
& & 5 & 8.083 & 496 & \multicolumn{3}{c}{Figures~3d and 3f at $\delta_c=0~\rm nm$} \\
& & 6 & 8.768 & 465 & \multicolumn{3}{c}{Figures~3d and 3f at $\delta_c=0~\rm nm$} \\
& & 7 & 9.236 & 443 & \multicolumn{3}{c}{Figures~3d and 3f at $\delta_c=0~\rm nm$} \\
& & 8 & 7.915 & 198 & \multicolumn{3}{c}{Figures~3d and 3f at $\delta_c=0~\rm nm$} \\
& & 9 & 8.198 & 495 & \multicolumn{3}{c}{Figures~3d and 3f at $\delta_c=0~\rm nm$} \\
& & 10 & 8.395 & 537 & \multicolumn{3}{c}{Figures~3d and 3f at $\delta_c=0~\rm nm$} \\
& & 11 & 8.888 & 481 & \multicolumn{3}{c}{Figures~3d and 3f at $\delta_c=0~\rm nm$} \\
& & 12 & 9.070 & 435 & \multicolumn{3}{c}{Figures~3d and 3f at $\delta_c=0~\rm nm$} \\
35 & 4.25 & 1 & 17.781 & 958 & \multicolumn{3}{c}{same sample as in Table~\ref{Table_1}} \\
& & 2 & 17.228 & 445 & \multicolumn{3}{c}{same sample as in Table~\ref{Table_1}} \\
& & 3 & 17.856 & 821 & \multicolumn{3}{c}{same sample as in Table~\ref{Table_1}} \\
& & 4 & 17.335 & 412 & \multicolumn{3}{c}{same sample as in Table~\ref{Table_1}} \\
& & 5 & 17.242 & 317 & \multicolumn{3}{c}{same sample as in Table~\ref{Table_1}} \\
& & 6 & 18.018 & 864 & \multicolumn{3}{c}{same sample as in Table~\ref{Table_1}} \\
& & 7 & 17.278 & 375 & \multicolumn{3}{c}{same sample as in Table~\ref{Table_1}} \\
& & 8 & 17.938 & 895 & \multicolumn{3}{c}{same sample as in Table~\ref{Table_1}} \\
& & 9 & 17.684 & 1023 & \multicolumn{3}{c}{same sample as in Table~\ref{Table_1}} \\
& & 10 & 17.717 & 1061 & \multicolumn{3}{c}{same sample as in Table~\ref{Table_1}} \\
& & 11 & 16.953 & 346 & \multicolumn{3}{c}{same sample as in Table~\ref{Table_1}} \\
& & 12 & 17.513 & 1367 & \multicolumn{3}{c}{same sample as in Table~\ref{Table_1}} \\
& & 13 & 17.548 & 1366 & \multicolumn{3}{c}{same sample as in Table~\ref{Table_1}} \\
& & 14 & 17.491 & 1300 & \multicolumn{3}{c}{same sample as in Table~\ref{Table_1}} \\
& & 15 & 1703 & 814 & \multicolumn{3}{c}{same sample as in Table~\ref{Table_1}} \\
& & 16 & 17.483 & 1356 & \multicolumn{3}{c}{same sample as in Table~\ref{Table_1}} \\
& & 17 & 17.437 & 1201 & \multicolumn{3}{c}{same sample as in Table~\ref{Table_1}} \\
& & 18 & 17.692 & 933 & \multicolumn{3}{c}{same sample as in Table~\ref{Table_1}} \\
& & 19 & 17.419 & 833 & \multicolumn{3}{c}{same sample as in Table~\ref{Table_1}} \\
& & 20 & 17.383 & 722 & \multicolumn{3}{c}{same sample as in Table~\ref{Table_1}} \\
& & 21 & 17.812 & 903 & \multicolumn{3}{c}{same sample as in Table~\ref{Table_1}} \\
& & 22 & 17.711 & 921 & \multicolumn{3}{c}{same sample as in Table~\ref{Table_1}} \\
& & 23 & 17.659 & 914 & \multicolumn{3}{c}{same sample as in Table~\ref{Table_1}} \\
& & 24 & 17.517 & 1036 & \multicolumn{3}{c}{same sample as in Table~\ref{Table_1}} \\
& & 25 & 17.450 & 1043 & \multicolumn{3}{c}{same sample as in Table~\ref{Table_1}} \\
& & 26 & 17.444 & 1107 & \multicolumn{3}{c}{same sample as in Table~\ref{Table_1}} \\
65 & 4.25 & 1 & 17.408 & 238 & \multicolumn{3}{c}{same sample as in Table~\ref{Table_1}} \\
& & 2 & 14.955 & 236 & \multicolumn{3}{c}{same sample as in Table~\ref{Table_1}} \\
& & 3 & 17.017 & 214 & \multicolumn{3}{c}{same sample as in Table~\ref{Table_1}} \\
65 & 2.75 & 1 & 32.855 & 226 & \multicolumn{3}{c}{same sample as in Table~\ref{Table_2}} \\
& & 2 & 27.884 & 250 & \multicolumn{3}{c}{same sample as in Table~\ref{Table_2}} \\
& & 3 & 27.087 & 243 & \multicolumn{3}{c}{same sample as in Table~\ref{Table_2}} \\
& & 4 & 28.080 & 255 & \multicolumn{3}{c}{same sample as in Table~\ref{Table_2}} \\

\hline

\end{tabularx}
\end{minipage}
\end{table}

\clearpage
\newpage

\bibliographystyle{apsrev4-1}
\bibliography{string}